\documentclass[twocolumn,aps,prl,showpacs,superscriptaddress]{revtex4}
\usepackage{psfig}
\newcommand {\snn}	{\sqrt{s_{_{\rm NN}}}}
\newcommand {\Npart}	{N_{\rm part}}
\newcommand {\Nmb}	{N_{\rm inc}}
\newcommand {\Bmb}	{B_{\rm inc}}
\newcommand {\Nbg}	{N_{\rm bg}}
\newcommand {\pt}	{p_{\perp}}
\newcommand {\kt}	{k_{\perp}}
\newcommand {\dphi}	{\Delta\phi}
\newcommand {\phitrig}	{\phi_{\rm t}}
\newcommand {\vtrig}	{v_2^{(t)}}
\newcommand {\vvtrig}	{v_4^{(t)}}
\newcommand {\STAR}	{}
\newcommand {\phenix}	{}

\begin{document}

%%%%%%%%%%%%%%%%%%%%%%%%%%%%%%%%%%%%%%%%%%%%%%%%%%%%%%%%%%%%%%%%%%%%%%

\title{Indications of Conical Emission of Charged Hadrons at the BNL Relativistic Heavy Ion Collider
%(version 10g, 1/22/2009, proof version from PRL)
}
\affiliation{Argonne National Laboratory, Argonne, Illinois 60439, USA}
\affiliation{University of Birmingham, Birmingham, United Kingdom}
\affiliation{Brookhaven National Laboratory, Upton, New York 11973, USA}
\affiliation{University of California, Berkeley, California 94720, USA}
\affiliation{University of California, Davis, California 95616, USA}
\affiliation{University of California, Los Angeles, California 90095, USA}
\affiliation{Universidade Estadual de Campinas, Sao Paulo, Brazil}
\affiliation{Carnegie Mellon University, Pittsburgh, Pennsylvania 15213, USA}
\affiliation{University of Illinois at Chicago, Chicago, Illinois 60607, USA}
\affiliation{Creighton University, Omaha, Nebraska 68178, USA}
\affiliation{Nuclear Physics Institute AS CR, 250 68 \v{R}e\v{z}/Prague, Czech Republic}
\affiliation{Laboratory for High Energy (JINR), Dubna, Russia}
\affiliation{Particle Physics Laboratory (JINR), Dubna, Russia}
\affiliation{Institute of Physics, Bhubaneswar 751005, India}
\affiliation{Indian Institute of Technology, Mumbai, India}
\affiliation{Indiana University, Bloomington, Indiana 47408, USA}
\affiliation{Institut de Recherches Subatomiques, Strasbourg, France}
\affiliation{University of Jammu, Jammu 180001, India}
\affiliation{Kent State University, Kent, Ohio 44242, USA}
\affiliation{University of Kentucky, Lexington, Kentucky, 40506-0055, USA}
\affiliation{Institute of Modern Physics, Lanzhou, China}
\affiliation{Lawrence Berkeley National Laboratory, Berkeley, California 94720, USA}
\affiliation{Massachusetts Institute of Technology, Cambridge, MA 02139-4307, USA}
\affiliation{Max-Planck-Institut f\"ur Physik, Munich, Germany}
\affiliation{Michigan State University, East Lansing, Michigan 48824, USA}
\affiliation{Moscow Engineering Physics Institute, Moscow Russia}
\affiliation{City College of New York, New York City, New York 10031, USA}
\affiliation{NIKHEF and Utrecht University, Amsterdam, The Netherlands}
\affiliation{Ohio State University, Columbus, Ohio 43210, USA}
\affiliation{Panjab University, Chandigarh 160014, India}
\affiliation{Pennsylvania State University, University Park, Pennsylvania 16802, USA}
\affiliation{Institute of High Energy Physics, Protvino, Russia}
\affiliation{Purdue University, West Lafayette, Indiana 47907, USA}
\affiliation{Pusan National University, Pusan, Republic of Korea}
\affiliation{University of Rajasthan, Jaipur 302004, India}
\affiliation{Rice University, Houston, Texas 77251, USA}
\affiliation{Universidade de Sao Paulo, Sao Paulo, Brazil}
\affiliation{University of Science \& Technology of China, Hefei 230026, China}
\affiliation{Shanghai Institute of Applied Physics, Shanghai 201800, China}
\affiliation{SUBATECH, Nantes, France}
\affiliation{Texas A\&M University, College Station, Texas 77843, USA}
\affiliation{University of Texas, Austin, Texas 78712, USA}
\affiliation{Tsinghua University, Beijing 100084, China}
\affiliation{United States Naval Academy, Annapolis, MD 21402, USA}
\affiliation{Valparaiso University, Valparaiso, Indiana 46383, USA}
\affiliation{Variable Energy Cyclotron Centre, Kolkata 700064, India}
\affiliation{Warsaw University of Technology, Warsaw, Poland}
\affiliation{University of Washington, Seattle, Washington 98195, USA}
\affiliation{Wayne State University, Detroit, Michigan 48201, USA}
\affiliation{Institute of Particle Physics, CCNU (HZNU), Wuhan 430079, China}
\affiliation{Yale University, New Haven, Connecticut 06520, USA}
\affiliation{University of Zagreb, Zagreb, HR-10002, Croatia}

\author{B.~I.~Abelev}\affiliation{University of Illinois at Chicago, Chicago, Illinois 60607, USA}
\author{M.~M.~Aggarwal}\affiliation{Panjab University, Chandigarh 160014, India}
\author{Z.~Ahammed}\affiliation{Variable Energy Cyclotron Centre, Kolkata 700064, India}
\author{B.~D.~Anderson}\affiliation{Kent State University, Kent, Ohio 44242, USA}
\author{D.~Arkhipkin}\affiliation{Particle Physics Laboratory (JINR), Dubna, Russia}
\author{G.~S.~Averichev}\affiliation{Laboratory for High Energy (JINR), Dubna, Russia}
\author{Y.~Bai}\affiliation{NIKHEF and Utrecht University, Amsterdam, The Netherlands}
\author{J.~Balewski}\affiliation{Massachusetts Institute of Technology, Cambridge, MA 02139-4307, USA}
\author{O.~Barannikova}\affiliation{University of Illinois at Chicago, Chicago, Illinois 60607, USA}
\author{L.~S.~Barnby}\affiliation{University of Birmingham, Birmingham, United Kingdom}
% ...

\collaboration{STAR Collaboration}\noaffiliation

\begin{abstract}
Three-particle azimuthal correlation measurements with a high transverse momentum trigger particle are reported for $pp$, d+Au, and Au+Au collisions at $\snn=200$~GeV by the STAR experiment. Dijet structures are observed in $pp$, d+Au and peripheral Au+Au collisions. An additional structure is observed in central Au+Au data, signaling conical emission of correlated charged hadrons. The conical emission angle is found to be $\theta$=$1.37\pm0.02{\rm(stat)}^{+0.06}_{-0.07}{\rm(syst)}$, independent of $\pt$.
\end{abstract}


\maketitle

%%%%%%%%%%%%%%%%%%%%%%%%%%%%%%%%%%%%%%%%%%%%%%%%%%%%%%%%%%%%%%%%%%%%%%
%{\em Introduction--} 
Collisions at BNL's Relativistic Heavy Ion Collider (RHIC) create a hot and dense medium that cannot be described by hadronic degrees of freedom~\cite{whitepapers}. Evidence of this is provided, in part, by jet-quenching: on the away side of a high transverse momentum ($\pt$) trigger particle (in azimuth relative to the trigger particle, $\dphi=\phi-\phitrig\approx\pi$), the correlated yield is strongly suppressed at $\pt>2$~GeV/$c$~\cite{b2b}, while at lower $\pt$ the yield is enhanced and the correlated hadrons appear to be partially equilibrated with the bulk medium and are broadly distributed in azimuth~\cite{jetspectra}. A number of physics mechanisms may account for the data: broadened jets due to large angle gluon radiation~\cite{Vitev}, deflected jets due to collective radial flow of the bulk~\cite{Armesto} or pathlength dependent energy loss~\cite{Hwa}, and conical emission due to $\check{\rm C}$erenkov gluon radiation~\cite{Cerenkov} or Mach-cone shock waves generated by large energy deposition in the hydrodynamic medium~\cite{conicalflow,Renk}. 

Identifying the underlying mechanism is important as it may probe the medium properties such as its speed of sound and equation of state~\cite{conicalflow,Renk}. To discriminate between the various mechanisms, we have performed an analysis of three-particle azimuthal correlations between a high $\pt$ trigger particle and two lower $\pt$ associated particles in $\dphi_i=\phi_i-\phitrig$ ($i$=1,2)~\cite{thesis}. We integrate over the pseudo-rapidity ($\eta$) direction because the near- and away-side jets are not correlated in $\eta$~\cite{jetspectra}. Many mechanisms predict that pairs of associated hadrons will be shifted away from $\dphi=\pi$, but will remain close to each other ($\dphi_1\approx\dphi_2$)~\cite{Vitev,Armesto,Hwa}. In contrast, the Mach-cone or $\check{\rm C}$erenkov radiation scenarios would result in particle emission on a cone around the away-side jet axis. When projected onto the azimuthal direction, the strongest signal of conical emission would be Jacobian peaks where pairs of correlated hadrons appear with equal probability to be close together or to be far apart and symmetric about $\pi$ (i.e., $\dphi_1-\pi\approx\pi-\dphi_2$)~\cite{Renk,method}. The latter feature is specific to conical emission. In this letter, we present evidence for correlated hadron pairs that are symmetrically located about $\pi$ relative to the trigger particle. The analysis is carried out with trigger and associated particles of $3<\pt<4$~GeV/$c$ and $1<\pt<2$~GeV/$c$, respectively, in $pp$, d+Au, and Au+Au collisions at $\snn=200$~GeV.

%%%%%%%%%%%%%%%%%%%%%%%%%%%%%%%%%%%%%%%%%%%%%%%%%%%%%%%%%%%%%%%%%%%%%%
%{\em Experiment--} 
Details of the STAR (Solenoidal Tracker at RHIC) experiment are described elsewhere~\cite{StarNIM}. This analysis uses $2$$\times$$10^6$ $pp$, $6.5$$\times$$10^6$ d+Au, and $1.2$$\times$$10^7$ minimum bias (MB) and $1.9$$\times$$10^7$ central trigger Au+Au events taken in 2001-2004. The central trigger data set corresponds to approximately 12\% of the total geometric cross-section, and will be henceforth referred to as ``12\% central" collisions. Charged particles are reconstructed with the Time Projection Chamber (TPC)~\cite{TPC}, which sits in a uniform 0.5~Tesla magnetic field. The Au+Au data are divided into nine centrality bins according to the charged particle multiplicity in the pseudorapidity region $|\eta|$$<$$0.5$ as in~\cite{specpaper}. Similarly the d+Au data are divided into three centrality bins of 0-10\%, 10-20\%, and 20-100\%. The trigger and associated particles are restricted to $|\eta|$$<$$1$. Our results are corrected for the centrality-, $\pt$-, and $\phi$-dependent reconstruction efficiency for associated particles and the $\phi$-dependent efficiency for trigger particles, and are normalized per corrected trigger particle.

%%%%%%%%%%%%%%%%%%%%%%%%%%%%%%%%%%%%%%%%%%%%%%%%%%%%%%%%%%%%%%%%%%%%%%
%{\em Analysis--}
Various approaches may be taken to measure three-particle correlations~\cite{thesis,method,cumulant}. This analysis treats the event as composed of two components: one is correlated with the trigger, $\hat{Y}_2$, and the other is background uncorrelated with the trigger except the indirect correlation via anisotropic flow. The correlated particle distribution (two-particle correlation) is thus given by
\begin{equation}
\hat{Y}_2(\dphi) = Y_2(\dphi) - a\Bmb F_2(\dphi)\, ,
\label{eq:twoparticle}
\end{equation}
where $Y_2(\dphi)=dN/d\dphi$ is the raw associated particle density per trigger. The other, background term is constructed by mixing triggers with different inclusive events (i.e. MB events within a given centrality bin), with the effect of anisotropic flow, 
\begin{equation}
F_2(\dphi) = 1 + 2\vtrig v_2\cos(2\dphi) + 2\vvtrig v_4\cos(4\dphi)\, ,
\label{eq:Ftwo}
\end{equation}
constructed pair-wise using flow measurements ($v_n^{\rm(t)}$ and $v_n$, $n=2,4$, are trigger and associated particle $n^{\rm th}$ harmonic coefficients, respectively)~\cite{v2paper,v4paper}; $\Bmb=\Nmb/2\pi$ is the inclusive event associated multiplicity density; $a=\Nbg/\Nmb$ scales $\Nmb$ to the underlying background associated multiplicity $\Nbg$ in trigger events, as discussed below.

In our {\it two-component approach}, the full three-particle distribution, $Y_3$, consists of the correlated triplets of interest, $\hat{Y}_3$, sets of three particles that are uncorrelated with each other except via flow, and cases where two of the particles are correlated (including jets and other correlations such as resonance decays) and the third is uncorrelated with the first two except via flow. The correlated pair distribution (three-particle correlation) is obtained by~\cite{thesis,method}
%\begin{eqnarray}
%\hat{Y}_3(\dphi_1,\dphi_2) &=& Y_3(\dphi_1,\dphi_2) \nonumber\\
%& & \hspace*{-0.6in} - a\Bmb\left[\hat{Y}_2(\dphi_1)F_2(\dphi_2)+\hat{Y}_2(\dphi_2)F_2(\dphi_1)\right] \nonumber\\
%& & \hspace*{-0.6in} - ba^2 Y_2^{\rm inc}(\dphi_1,\dphi_2)\left[1+\frac{F_3(\dphi_1,\dphi_2)}{F_2(\dphi_1-\dphi_2)}\right], \label{eq:threeparticle}
%\end{eqnarray}
\begin{widetext}
\begin{eqnarray}
\hat{Y}_3(\dphi_1,\dphi_2) &=& Y_3(\dphi_1,\dphi_2) - a\Bmb\left[\hat{Y}_2(\dphi_1)F_2(\dphi_2)+\hat{Y}_2(\dphi_2)F_2(\dphi_1)\right] \nonumber\\
& & - ba^2 Y_2^{\rm inc}(\dphi_1,\dphi_2)\left[1+\frac{F_3(\dphi_1,\dphi_2)}{F_2(\dphi_1-\dphi_2)}\right], \label{eq:threeparticle}
\end{eqnarray}
\end{widetext}
where $Y_3=d^2 N/d\dphi_1d\dphi_2$ is the raw associated particle pair density per trigger, and the second and third terms on the r.h.s.~are backgrounds. The second term, referred to as {\em trig-corr-bkgd}, arises from combining a correlated trigger-associated pair with a background particle, and is constructed from the product of the two-particle correlation and its flow modulated background.

The third term, referred to as {\em trig-bkgd-bkgd}, arises from combining a trigger with two background particles, and contains all correlations between the two background particles as well as the flow correlation between them and the trigger. The former is the inclusive event pair density $Y_2^{\rm inc}=d^2\Nmb/d\dphi_1d\dphi_2$ relative to a random trigger $\phitrig$, which is constructed by mixing the trigger from one event with two particles from another, inclusive event. The latter is referred to as {\em trigger flow}, where 
\begin{eqnarray}
F_3(\dphi_1,\dphi_2) 
&=& F_2(\dphi_1)+F_2(\dphi_2) - 2 \nonumber \\
&+& 2\vtrig v_2^{(1)}v_4^{(2)}\cos 2(\dphi_1-2\dphi_2) \nonumber \\
&+& 2\vtrig v_2^{(2)}v_4^{(1)}\cos 2(2\dphi_1-\dphi_2) \nonumber \\
&+& 2v_2^{(1)}v_2^{(2)}\vvtrig\cos 2(\dphi_1+\dphi_2)
\label{eq:Fthree}
\end{eqnarray}
is constructed triplet-wise by mixing the trigger with particles from two different inclusive events. The factor $ba^2$ scales the number of pairs in inclusive events, $\langle\Nmb(\Nmb-1)\rangle$, to that in the underlying background, $\langle\Nbg(\Nbg-1)\rangle$. Non-Poisson multiplicity fluctuations, which can be different in inclusive events and in the background underlying trigger events, result in deviations of $b$ from one. We approximate $b$ by $\frac{\langle N(N-1)\rangle/\langle N\rangle^2}{\langle\Nmb(\Nmb-1)\rangle/\langle\Nmb\rangle^2}$ where $N$ is the associated multiplicity in trigger events. 

The analysis procedure is performed and the scaling factors $a$ and $b$ are determined for each centrality bin separately; the final three-particle results are combined over centrality bins to increase the statistics. The value of $a$ is determined assuming that the three-particle correlation signal has Zero Yield at Minimum (3-ZYAM); the total size of the minimum signal regions is chosen to be 10\% of $(2\pi)^2$. It is so chosen so that it is small enough to approximate the real minimum, but large enough to avoid large statistical fluctuations. This size is varied between 5-15\% of $(2\pi)^2$, keeping $a$ fixed, to assess the systematic uncertainty on $b$. The upper end of the systematic uncertainty on $a$ is taken to be the $a$ value from two-particle ZYA1 (Zero Yield At 1 radian) where $\hat{Y}_2(\dphi)$ vanishes at $|\dphi\pm 1|<\pi/18$~\cite{jetspectra}. The lower end is determined, while keeping $b$ at its default value, from the lowest data point (out of total 24$\times$24), which should be lower than the true 3-ZYAM because of statistical fluctuations. With $a$ at each systematic end, the value of $b$ is readjusted, shifting the three-particle correlation result by an approximately constant pedestal, to preserve 3-ZYAM. For the top 5\% centrality fraction with the 12\% central data, $a=0.994(^{+0.005}_{-0.004})$ and $b=1.00021(^{+0.00003}_{-0.00005})$.

\begin{figure*}[hbt]
\centerline{\psfig{file=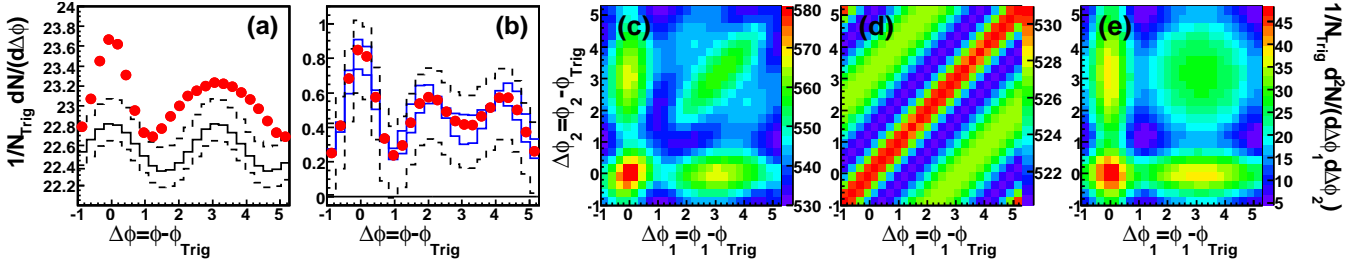,width=\textwidth}}
\caption{(a) Raw two-particle correlation signal $Y_2$ (red), background $a\Bmb F_2$ (solid histogram), and background systematic uncertainty from $a$ (dashed histograms). (b) Background-subtracted two-particle correlation $\hat{Y}_2$ (red), and systematic uncertainties due to $a$ (dashed histograms) and flow (blue histograms). (c) Raw three-particle correlation $Y_3$. (d) $ba^2 Y_2^{\rm inc}$. (e) Sum of trig-corr-bkgd and trigger flow. Data are from 12\% central Au+Au collisions. Statistical errors in (a,b) are smaller than the point size.}
\label{fig:analysis}
\end{figure*}

Figure~\ref{fig:analysis} shows two-particle correlations in Au+Au central collisions: the raw $Y_2(\dphi)$ and the $a$-scaled background $a\Bmb F_2(\dphi)$ in (a), and the background subtracted $\hat{Y}_2(\dphi)$ in (b). Fitting $\hat{Y}_2(\dphi)$ to various functional forms similar to those in Ref.~\cite{phenix} yields away-side peaks centered 1.18-1.34 radians from $\pi$. Figure~\ref{fig:analysis}(c,d,e) show, respectively, the raw three-particle correlation $Y_3(\dphi_1,\dphi_2)$, $ba^2 Y_2^{\rm inc}$, and the trig-corr-bkgd term plus trigger flow~\cite{note_symm}. 

%{\em Systematic uncertainties--}
Table~\ref{tab:sys} summarizes the major sources of systematic uncertainties. 
(I) Uncertainty in the normalization factor $a$ is assessed as above.
(II) The $v_2$ used is the average of modified reaction plane $v_2$\{{\sc mrp}\} and 4-particle cumulant $v_2$\{4\}~\cite{jetspectra}. The two-particle cumulant $v_2$\{2\}, which contains flow fluctuations and potentially non-flow effects, and the $v_2$\{4\} or $v_2$\{{\sc 2d}\} (obtained from a two-dimensional analysis in $\Delta\eta$ and $\Delta\phi$) bracket the systematic uncertainties. We used a parameterization of $v_4=1.15\,v_2^2$~\cite{v4paper} and the $v_2$ uncertainties are propagated.
(III) The trig-corr-bkgd term in Eq.(\ref{eq:threeparticle}) is constructed from the two-particle correlation and its background, both averaged over the reaction plane (RP) angle. The effect of the change of the correlation structure with the angle between the trigger and the RP~\cite{RPjetCorr} is estimated and included in our final results. The size of the effect is assigned as a single-sided systematic uncertainty. The systematic uncertainty from (I) primarily impacts the overall magnitude of the correlation, with little influence on the shape, whereas those from (II) and (III) have a smaller impact on the magnitude, but affect the shape of the correlation. 

Table~\ref{tab:sys} also lists the total systematic uncertainty from other, minor sources: 
uncertainty in the normalization factor $b$ estimated as above; 
$\pm 20$\% uncertainty on the unmeasured $\vvtrig$~\cite{v4paper};
uncertainties due to the finite centrality bins on trig-corr-bkgd and trig-bkgd-bkgd terms estimated by breaking each centrality into finer bins;
and 10\% uncertainty due to the efficiency correction.

\begin{table}[hbt]
\caption{Systematic uncertainties on three-particle correlation strength on the away side: central region ($|\dphi_{1,2}-\pi|<0.35$) and off-diagonal region ($|\dphi_1-\pi\pm1.37|<0.35$ and $|\dphi_2-\pi\mp1.37|<0.35$).}
\label{tab:sys}
\begin{ruledtabular}
\begin{tabular}{l|c|cc|cc}
source & d+Au & \multicolumn{2}{c|}{50-30\% Au+Au} & \multicolumn{2}{c}{12\% central Au+Au} \\
& cent. & cent. & off-diag. & cent. & off-diag. \\ \hline
(I) $a$ & $^{+16}_{-18}$\% 
	& $^{+29}_{-60}$\% & $^{+30}_{-63}$\% 
	& $^{+42}_{-61}$\% & $^{+21}_{-32}$\% \\
(II) $v_2$	& --
		& $^{-8}_{+17}$\% & $^{+36}_{-14}$\% 
		& $^{-13}_{+45}$\% & $^{+32}_{-16}$\% \\
(III) RP	& -- & $+11$\% & $+8$\% & $+32$\% & $+5$\% \\ 
others	 	& $^{+11}_{-10}$\% 
		& $^{+12}_{-11}$\% & $^{+18}_{-12}$\% 
		& $^{+16}_{-20}$\% & $\pm 12$\% \\
\end{tabular}
\end{ruledtabular}
\end{table}

%%%%%%%%%%%%%%%%%%%%%%%%%%%%%%%%%%%%%%%%%%%%%%%%%%%%%%%%%%%%%%%%%%%%%%
\begin{figure}[hbt]
\centerline{\psfig{file=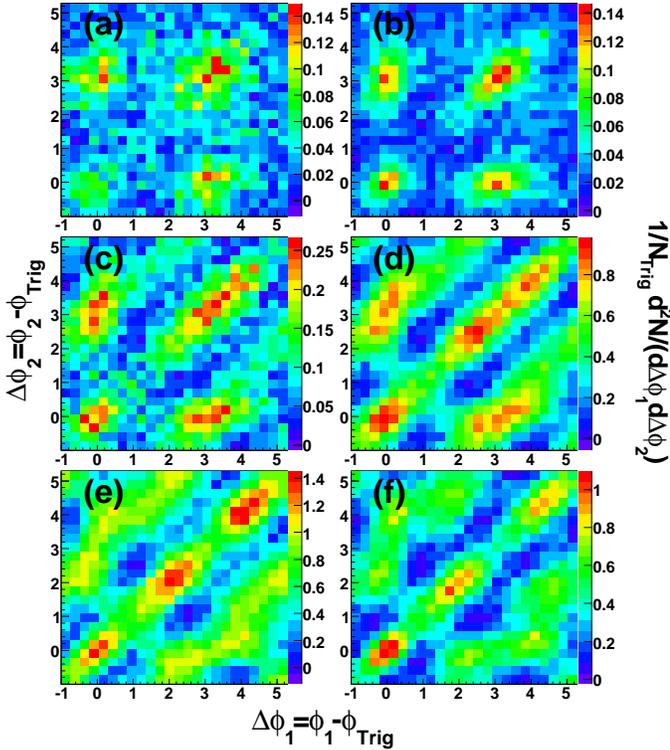,width=0.5\textwidth}}
\caption{Background subtracted three-particle correlations, $\hat{Y}_3$, for (a) $pp$, (b) d+Au, (c) 80-50\% Au+Au, (d) 50-30\% Au+Au, (e) 30-10\% Au+Au, and (f) 12\% central Au+Au. Statistical errors per bin are approximately $\pm0.012$ in (a) and $\pm0.006$ in (b), both at $(\pi,\pi)$, and are $\pm0.022$, $\pm0.049$, $\pm0.099$ and $\pm0.077$ from (c) to (f), similar for all bins.}
\label{fig:results}
\end{figure}

%{\em Results and discussion--}
Figure~\ref{fig:results} shows the background subtracted three-particle azimuthal correlations, $\hat{Y}_3$, in MB $pp$, d+Au, and three combined centrality bins of MB Au+Au and the 12\% central collisions. Four distinct peaks are observed for each data set, corresponding to both correlated particles on the near side (around $\dphi_1=\dphi_2=0$), both on the away side (around $\pi$), and one on each side. The near-side peaks are slightly elongated along the diagonal, probably due to momentum balance in combination with the fact that the trigger direction differs from its parent's.

\begin{figure}[hbt]
\centerline{\psfig{file=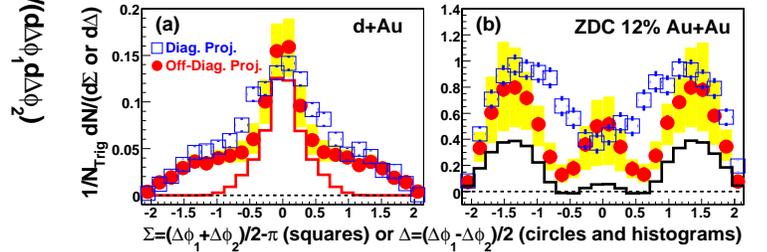,width=0.5\textwidth}}
\caption{Projections of away-side three-particle correlations along the diagonal $\Sigma$ within $0<\Delta<0.35$ (squares) and along the off-diagonal $\Delta$ within $|\Sigma|<0.35$ (points) in (a) d+Au and (b) 12\% central Au+Au collisions. The shaded areas indicate systematic uncertainties on the off-diagonal projections. The histogram in (a) is the near-side off-diagonal projection. The histogram in (b) is the away-side off-diagonal projection of our result with $a=b=1$.}
\label{fig:projections}
\end{figure}

The away-side central peak is elongated along the diagonal, progressively from $pp$ to d+Au to Au+Au collisions. This indicates that the away-side pairs stay relatively close while their angles vary over a wide range event-by-event. Figure~\ref{fig:projections}(a) shows the effect quantitatively by projecting the d+Au three-particle correlation on the away side ($1<\dphi_{1,2}<2\pi-1$) along the diagonal in $\Sigma=(\dphi_1+\dphi_2)/2-\pi$ and off-diagonal in $\Delta=(\dphi_1-\dphi_2)/2$, within the ranges of $0<\Delta<0.35$ and $|\Sigma|<0.35$, respectively~\cite{note_symm}. For comparison the off-diagonal projection on the near side ($|\dphi_{1,2}|<1$) is also shown. 

For central Au+Au collisions, additional peaks are observed in Fig.~\ref{fig:results} on the away side along the off-diagonal, indicating large opening angles between the away-side correlated pairs, symmetric about $\pi$, $\dphi_1-\pi\approx\pi-\dphi_2$ corresponding to each off-diagonal peak. The observed correlation pattern in central collisions is quite different from the expectations for statistical global momentum conservation~\cite{Borghini}. Figure~\ref{fig:projections}(b) shows the diagonal and off-diagonal projections of the away-side three-particle correlation result from the 12\% central data. The off-diagonal projection of our result with $a=b=1$ is also shown. The off-diagonal side peaks are prominent; these peaks are evidence of conical emission of charged hadrons correlated with high $\pt$ trigger particles. The side peaks in the diagonal projection contain conical emission and possibly other contributions such as $\kt$ broadening, large angle gluon radiation, and deflected jets.

The angular distance $\theta$ of the off-diagonal peak locations from $\pi$ is obtained by fitting the off-diagonal projections to a central plus two symmetric side Gaussians. For 12\% central Au+Au, $\theta$=1.37$\pm$0.02(stat)$^{+0.06}_{-0.07}$(syst)~radian. The difference between $\theta$ and the fit position to two-particle correlation may arise because the latter measures a combination of effects. The value of $\theta$ does not depend on centrality or the associated particle $\pt$. For $\pt$=0.5-1, 1-1.5, 1.5-2, 2-3~GeV/$c$, $\theta$=1.38$\pm$0.03(stat)$^{+0.07}_{-0.05}$(syst), 1.36$\pm$0.04$^{+0.08}_{-0.07}$, 1.29$\pm$0.04$^{+0.19}_{-0.10}$, and 1.31$\pm$0.05$^{+0.25}_{-0.09}$, respectively. If the observed conical emission is generated by Mach-cone shock waves, the measured angle $\theta$ reflects the speed of sound in the created medium averaged over the evolution of the collision~\cite{conicalflow,Renk}.

\begin{figure}[hbt]
\centerline{\psfig{file=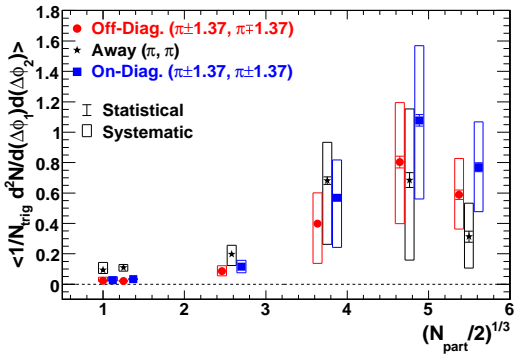,width=0.4\textwidth}}
\caption{Three-particle correlation strength per radian$^2$ versus $(\Npart/2)^{1/3}$ where $\Npart$ is the number of participants. Some of the data points have been displaced in $N_{\rm part}$ for clarity.}
\label{fig:strength}
\end{figure}

To characterize the correlation strength, the average signals are evaluated within 0.7$\times$0.7 radian$^2$ centered at $(\dphi_1,\dphi_2)=(\pi,\pi)$, $(\pi\pm1.37,\pi\pm1.37)$, and $(\pi\pm1.37,\pi\mp1.37)$. Figure~\ref{fig:strength} shows the average signal strength~\cite{note_symm} in $pp$, d+Au, and Au+Au collisions as a function of $(\Npart/2)^{1/3}$. The signal strength increases and appears to saturate in central collisions. While the away central peak is the dominant structure in $pp$, d+Au, and peripheral Au+Au, the diagonal and off-diagonal side peaks increase rapidly in strength with centrality and become the dominant structures in central Au+Au collisions.

%%%%%%%%%%%%%%%%%%%%%%%%%%%%%%%%%%%%%%%%%%%%%%%%%%%%%%%%%%%%%%%%%%%%%%
%{\em Conclusions--}
In conclusion, the first three-particle azimuthal correlation measurements with a high transverse momentum trigger particle are reported by the STAR experiment. The analysis treats the event as composed of two components, one correlated with the trigger and the other, background. Results are presented for minimum bias $pp$, d+Au, and different centralities in Au+Au collisions at $\snn=200$~GeV between a trigger particle of $3<\pt<4$~GeV/$c$ and two associated particles of $1<\pt<2$~GeV/$c$. Dijet structures are observed in $pp$, d+Au and peripheral Au+Au collisions, with a progressive diagonal elogation of the away-side central peak. Distinct peaks at $\theta$=$1.37\pm0.02{\rm(stat)}^{+0.06}_{-0.07}{\rm(syst)}$ from $\pi$ are observed on the away side in central Au+Au collisions, with correlated hadron pairs far apart, symmetric about $\pi$. These structures are evidence of conical emission of hadrons correlated with high $\pt$ particles. The conical emission angle is measured to be independent of the associated particle $\pt$.

%%%%%%%%%%%%%%%%%%%%%%%%%%%%%%%%%%%%%%%%%%%%%%%%%%%%%%%%%%%%%%%%%%%%%%
%{\em Acknowledgments--}
We thank the RHIC Operations Group and RCF at BNL, and the NERSC Center at LBNL and the resources provided by the Open Science Grid consortium for their support. This work was supported in part by the Offices of NP and HEP within the U.S.~DOE Office of Science, the U.S.~NSF, the Sloan Foundation, the DFG Excellence Cluster EXC153 of Germany, CNRS/IN2P3, RA, RPL, and EMN of France, STFC and EPSRC of the United Kingdom, FAPESP of Brazil, the Russian Ministry of Sci.~and Tech., the NNSFC, CAS, MoST, and MoE of China, IRP and GA of the Czech Republic, FOM of the Netherlands, DAE, DST, and CSIR of the Government of India, Swiss NSF, the Polish State Committee for Scientific Research,  and the Korea Sci.~\& Eng.~Foundation.

%%%%%%%%%%%%%%%%%%%%%%%%%%%%%%%%%%%%%%%%%%%%%%%%%%%%%%%%%%%%%%%%%%%%%%


\begin{thebibliography}{99}
\bibitem{note_symm}
All three-particle results are symmeterized between $\dphi_1$ and $\dphi_2$. In the off-diagonal projections in Fig.~\ref{fig:projections} only half the data points are statistically independent; the slight asymmetry in the plots is due to binning. The statistical errors are multiplied by $\sqrt{2}$ in Fig.~\ref{fig:strength} to account for this.
\end{thebibliography}
\end{document}